\documentclass[aps,prl,nofootinbib,twocolumn,floatfix,letterpaper,superscriptaddress,showpacs]{revtex4}
\usepackage{graphicx}\usepackage{natbib}\usepackage{amsmath}
\pdfoutput=1
\begin{document}

\title{Galactic Streams of Cosmic-ray Electrons and Positrons}

\author{Matthew D. Kistler}
\affiliation{Lawrence Berkeley National Laboratory and Department of Physics, University of California, Berkeley, CA 94720}
\affiliation{Einstein Fellow}

\author{Hasan Y{\"u}ksel}
\affiliation{Theoretical Division, Los Alamos National Laboratory, Los Alamos, NM 87544}

\author{Alexander Friedland}
\affiliation{Theoretical Division, Los Alamos National Laboratory, Los Alamos, NM 87544}

%
%

\begin{abstract}
Isotropic diffusion is a key assumption in many models of cosmic-ray electrons and positrons.  We find that simulation results imply a critical energy of $\sim\,$10--1000~GeV above which $e^\pm$ can spend their entire lives in streams threading magnetic fields, due to energy losses.  This would restrict the number of $e^\pm$ sources contributing at Earth, likely leading to smooth $e^\pm$ spectra, as is observed.  For positrons, this could be as few as one, with an enhanced flux that would ease energetics concerns of a pulsar origin of the positron excess, or even zero, bringing dark matter to the fore.  We conclude that ideas about $e^\pm$ propagation must be revised and discuss implications for recent AMS-02 data.
\end{abstract}


%
\maketitle

{\bf Introduction.---}
Energy losses due to synchrotron radiation and inverse-Compton scattering severely limit the distances over which cosmic-ray electrons and positrons can remain highly energetic \cite{Ginzburg(1964),Longair2010}.  Thus, observations of these particles provide an opportunity to examine the recent high-energy history of within a few hundred parsecs of Earth.  A number of experiments have contributed to remarkable developments in this area \cite{Adriani:2008zr,Aharonian:2008aaa,Fermi:2009zk,Adriani:2011xv,FermiLAT:2011ab,Aguilar:2013qda,Adriani:2013uda} and more are poised to break into the multi-TeV regime \cite{Aguilar:2002ad,Nutter,Tamura:2010zzc}.  The objects that may yield measurable fluxes in this range are few \cite{Kistler:2009wm} and represent perhaps the most immediate hope of directly discovering an extrasolar cosmic-ray source.

Making use of this data requires an understanding of the propagation of cosmic rays through magnetic fields after escaping their sources.  Many standard treatments for electrons and positrons (e.g., Refs.~\cite{Atoyan:1995,Hooper:2008kg,Yuksel:2008rf,Profumo:2008ms,Grasso:2009ma}) assume isotropic diffusion.  But why?  Particles care about two length scales of the field: the largest present, which funnels them in specific directions, and scales of order their Larmor radius, which result in scattering \cite{Jokipii(1966)}.  As we will see, the small scale turbulent variation in the field alone ends up being too weak to lead to isotropic diffusion for physically sensible assumptions in scenarios involving $e^\pm$.

\begin{figure}[b!]
\includegraphics[width=0.96\columnwidth,clip=true]{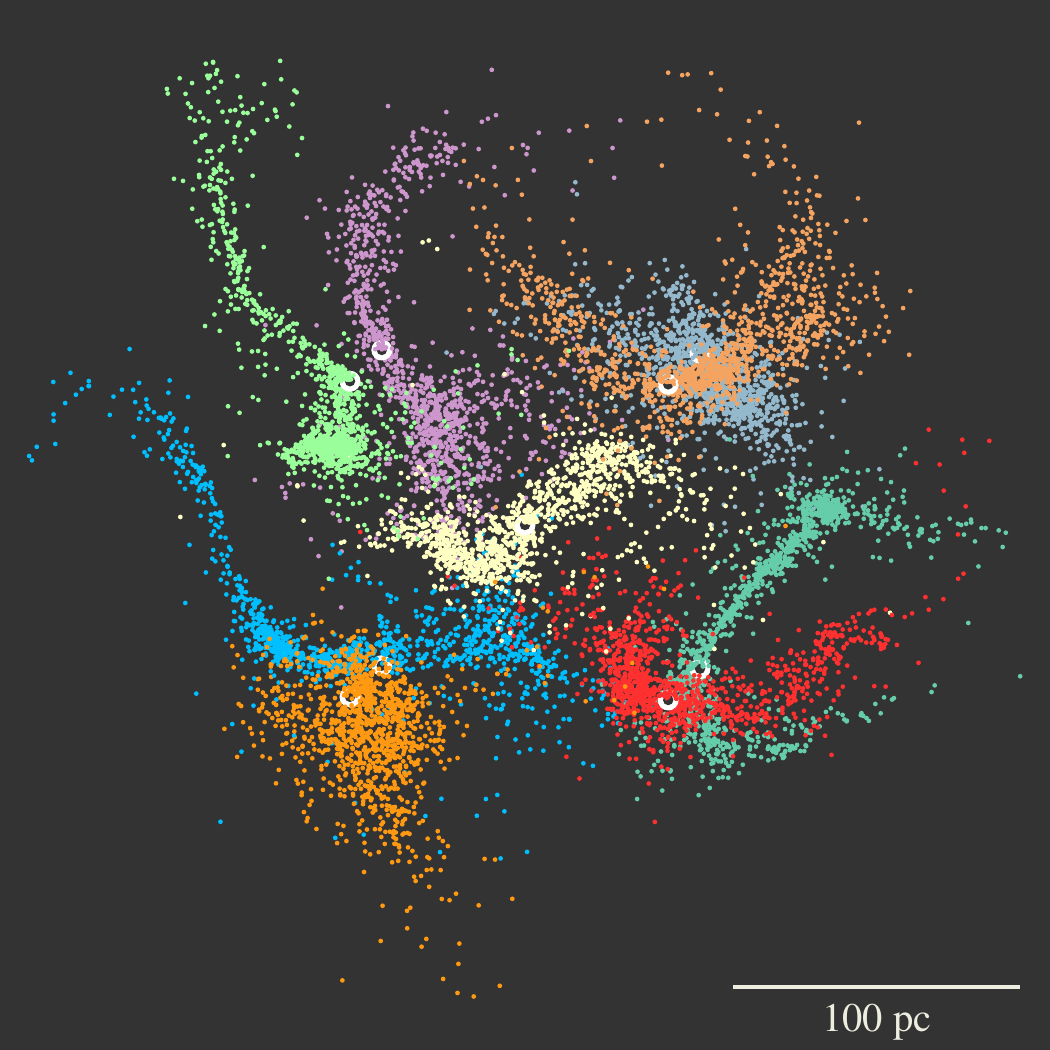}
\caption{Distributions of electrons with $1\,$PeV initial energies after propagating 5000~yr in a 3$\,\mu$G random magnetic field.  Each stream ({\it dots}) arises from  one of nine sources ({\it circles}).
\label{fig:star}}
\end{figure}

We compare simulations of $e^\pm$ propagation using random magnetic fields with such analytical solutions over a range of energies, distances, and times.  We concentrate on particular realizations of fields, rather than the average over many such fields as implicitly assumed in diffusion models, since we can observe particles directly from only one location in our Galaxy.  As an illustration, we show in Fig.~\ref{fig:star} particle trajectories arising from nine simulated $e^\pm$ sources distributed within one random field, as described in detail below.  We see, similar to \cite{Giacinti:2012ar}, that \textit{rather than being distributed in spherically symmetric distributions, particles are initially confined to filamentary structures}, which we refer to as ``\textit{streams}''.

Importantly, we take into account particle energy losses, which is crucial here, \textit{since cooling necessitates that electrons and positrons can only travel for a limited time with energies exceeding a given value}, in stark contrast to the case of protons.  We argue that, because of the losses, the consequence of this approach is that electrons and positrons with measured energies $\gtrsim\!10\!-\!1000$~GeV can be expected to have spent their entire lives within streams, {\it never reaching the isotropic diffusive regime}.

This containment of cosmic-ray electrons and positrons would be far reaching, including a need to abandon isotropic $e^\pm$ diffusion calculations in this regime.  We discuss how this limits the number of sources that can contribute to the measured $e^\pm$ spectra, and thus an absence of spectral features as would be expected from a multitude of potential nearby sources, in accord with the smooth $e^-\!/e^+$ spectra seen by PAMELA, Fermi, and AMS.  We address how this impacts pulsar models of the positron excess, $e^\pm$ anisotropies, cosmic-ray-induced turbulence, and gamma-ray ``halos'' from high-energy $e^\pm$.

{\bf Fields and streams.---}
To study cosmic-ray electrons traversing a number of magnetic field realizations, we simulate the propagation of large numbers of particles to obtain particle densities over a range of times and distances.  The Galactic field consists of a regular component with reversals in direction between spiral arms and a turbulent component with coherence length $\sim\,$100~pc \cite{Pshirkov:2011um}.  The dominant field surrounding the Sun is tilted \cite{Opher(2009)}, with the IBEX Ribbon implying a $\sim\!3\,\mu$G field oriented $\sim\!55^\circ$ off of the Galactic plane \cite{Funsten(2009),Frisch(2012)}.  Fluctuations consistent with Kolmogorov turbulence are inferred over a wide range of scales \cite{Minter(1996)}.  Radio data imply a local $\sim\!7.5\,\mu$G random field \cite{Strong:2011wd}, and a few $\mu$G turbulent field may explain small-scale PeV proton anisotropies \cite{Giacinti:2011mz}.

For a singly-charged particle within a magnetic field of strength $B$, the equation of motion can be written as
\begin{equation}
  \frac{d \, {\boldsymbol \beta}}{d\,t} \simeq 0.925 \, \frac{{\boldsymbol  \beta}\times {\bf B}}{E} \hspace*{1cm} {\boldsymbol  \beta} = \frac{d \,{\bf r}}{d\,t}\,,
\label{eq:motion}
\end{equation}
where velocity is given in terms of $c$ so that ${\boldsymbol \beta}$ is a unit vector.  Distance and time are expressed in pc, energy in PeV, and $B$ in $\mu$G, with Larmor radius $r_L\!\simeq\!1.08\,E/B$.

We first follow the formulation of Ref.~\cite{Yuksel:2012ee} (see also \cite{Tribble:1991tg,Murgia:2004zn,Casse:2001be,DeMarco:2007eh}) to obtain a $\sim\!3\,\mu$G magnetic field ${\bf B}_{k_0}^{k_1}$ between scales $k_0$ and $k_1$ whose power follows $\left| B_{k}\right|^2\!\propto\!k^{-(n+2)}$ ($n\!=\!5/3$ for Kolmogorov turbulence), addressing other possibilities later.  It is a practical impossibility to construct a magnetic field to the $k$ values associated with the $r_L$ of GeV-TeV particles, due to memory constraints.  We instead reach smaller scales by using nested boxes, ${\bf B}_{k_0}^{k_N}({\bf r})\!\propto\!\sum_{i=0}^{N}\eta^{-i/2}\,{\bf B}_{k_0}^{k_1}(\eta^i {\bf r})$, where $\eta\!=\!k_0/k_1$, that repeat periodically (similar to Ref.~\cite{Giacinti:2011ww}).  This maintains the proper normalization of power over all scales.

A sufficiently large $k_0$ value ensures variation in the simulation box even in the largest scales and $\eta$ can be as large as memory allows.  We normalize the rms value of $B_{k_0}^{k_N}$ to $3\,\mu$G and choose a maximum length scale, $l_{max}\!\propto\!1/k_0$, of 200~pc, the size of structures such as the Local Bubble \cite{Welsh:2009in,Frisch(2011)}, and a minimum scale, $l_{min}\!\propto\!1/k_N$, smaller than $r_L$, with a coherence length $l_c\!\simeq\!40\,$pc.  Particles are injected isotropically at the source and Eq.~(\ref{eq:motion}) is solved via a fourth-order Runge-Kutta method with step sizes smaller by at least a factor of several than $r_L$.

At the energies of interest here, electron energy losses arise from synchrotron radiation and inverse-Compton (IC) scattering on ambient photons.  For later comparisons, we approximate losses as continuous (IC is better treated as stochastic at very high energies \cite{Blumenthal:1970gc}), with $-dE/dt\!=\!b(E)\!=\!b_0\,E^2$, so $1/E\!=\!1/E_g\!+\!b_0 t$, where $E_g$ is the energy at generation.  We assume a $\sim\!3\,\mu$G magnetic field and the cosmic microwave background (CMB) alone here (since other backgrounds are suppressed; see Fig.~2 in \cite{Delahaye:2010ji}), so $b_0\!\simeq\!5\!\times\!10^{-17}$~s$^{-1}$~GeV$^{-1}$. 

In Fig.~\ref{fig:star}, we show the positions of cosmic-ray $e^\pm$ with $E_g\!=\!1\,$PeV after propagating for 5000~yr, a time $\sim\!10\times$ greater than their energy loss timescale ($t_l\!\sim\!(b_0\,E)^{-1}$), from sources at nine locations distributed within the same 3$\,\mu$G field configuration.  We see that, \textit{rather than diffusing isotropically, the fluxes are shaped by the large scale components of the local field}.  This has been seen using similar fields with Galactic proton sources \cite{Giacinti:2012ar}.

\begin{figure}[t!]
\includegraphics[width=0.99\columnwidth,clip=true]{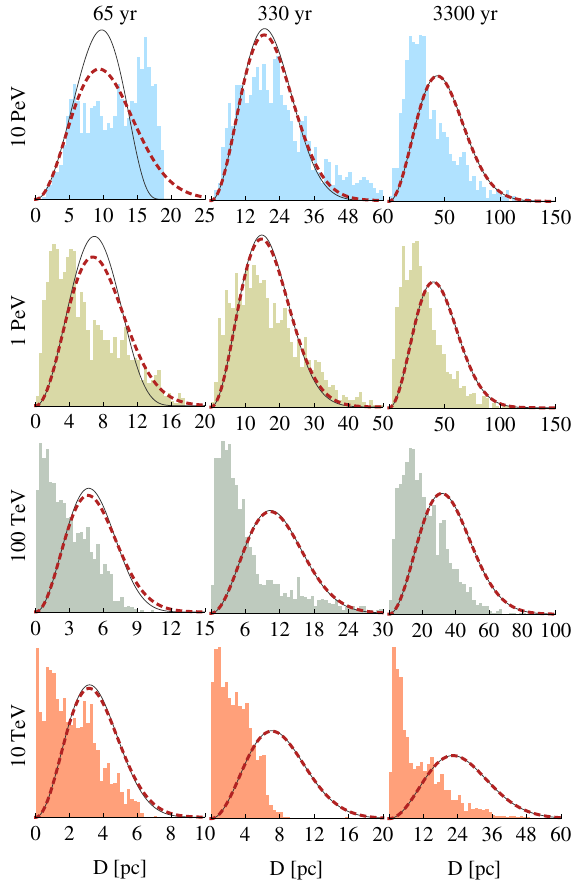}
\caption{Radial distributions of cosmic-ray $e^\pm$ with initial energies of 10~TeV--10~PeV after propagation through a 3$\,\mu$G random magnetic field at given times ({\it bins}).  Also shown are the expectations from isotropic diffusion ({\it dashed}) and J\"{u}ttner ({\it solid}) models, which are seen to poorly represent the data.
\label{fig:sky}}
\end{figure}

Using the source position located at the center of Fig.~\ref{fig:star}, we consider instantaneous bursts for $E_g$ covering 10~TeV--10~PeV (variable sources can be built up as combinations of successive bursts).  We display in Fig.~\ref{fig:sky} snapshots of the radial distributions summed over all directions for these four energies at three given times.

\begin{figure*}[t!]
\includegraphics[width= 1.8 \columnwidth,clip=true]{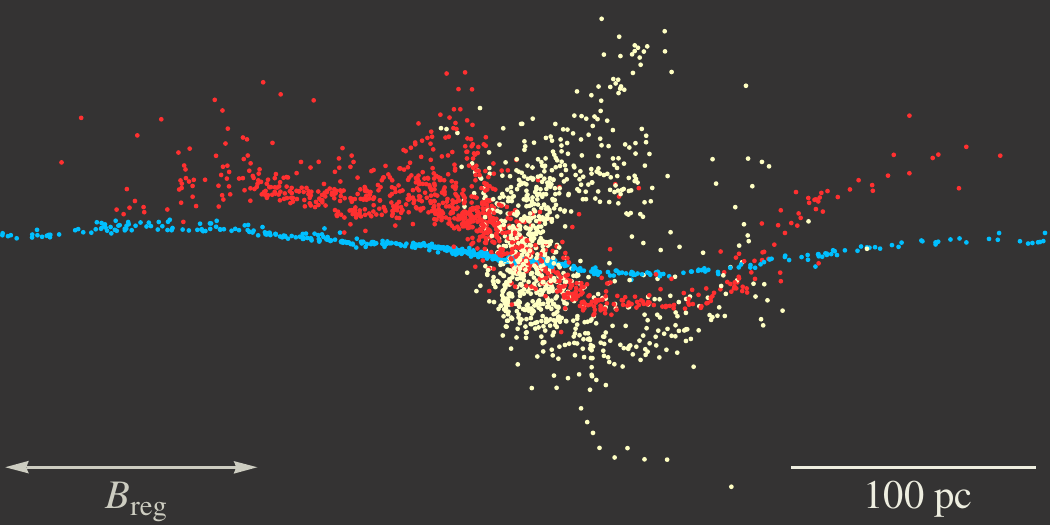}
\caption{Electron streams arising after propagating for 5000~yr with initial energies of $1\,$PeV (from a location at figure center).  Here, we vary the ratio of regular to random field magnitudes as 0 ({\it white}), 1 ({\it red}), and 5 ({\it blue}), while fixing $B_{\rm reg}\!+\!B_{\rm rand}\!=\!3\,\mu$G.  We see that increasing $B_{\rm reg}$ orients propagation along the regular field direction.
\label{fig:reg3}}
\end{figure*}

{\bf Comparison to analytical models.---}
Assuming spherical symmetry, the diffusion equation yields a particle density $n_d(r,t,E)$ from a bursting source \cite{Ginzburg(1964)} as
\begin{equation}
  n_d(r,t,E)= \frac{e^{-r^2/r_{\rm dif}^2}} {\pi^{3/2} \, r_{\rm dif}^3} \frac{dN}{dE_g}\frac{dE_g}{dE}
\,,
\label{eq:syrovatsky}
\end{equation}
with $\lambda(E,t)\!=$$\int_0^t dt' D[E(t')]\!=$$\int_E^{E_g} dE' D(E')/b(E')\,$giving $r_{\rm dif}(E,t)\!=\!2\sqrt{\lambda(E,t)}$.  $E_g$ is mapped to the measured $E$ after losses via $dE_g/dE\!=\!(E_g/E)^2$, using the above $b_0$.

Since the diffusive solution has been noted to fail at high energies \cite{Aloisio:2008tx,Kistler:2009wm}, we also consider a model based on the J\"{u}ttner particle distribution, which explicitly constrains $v<c$ to eliminate superluminal behavior, giving
\begin{equation}
  n_J(r,t,E) =  \frac{\theta[1-\xi] \, \alpha_J}{4 \pi (c t)^3 K_1(\alpha_J)}
           \frac{e^{-\alpha_J/\sqrt{1-\xi^2}}}{(1-\xi^2)^2}
            \frac{dN}{dE_g}\frac{dE_g}{dE}
,
\label{eq:juttner}
\end{equation}
where $\xi(r,t)$$\,=\,$$r/ct$, $\theta$ is the step function, $K_1$ is the modified Bessel function, and $\alpha_J(E,t)$$\,=\,$${c^2t^2}/({2\lambda(E,t)})$.

In Fig.~\ref{fig:sky}, we compare the simulated densities to both models, using $D(E)\!=\!5.4\!\times\!10^{26}\,(E/$GeV$)^{1/3}\,$cm$^2\,$s$^{-1}$, a value close to that seen in Ref.~\cite{Giacinti:2012ar}.  We note that this is \textit{lower by about an order of magnitude than used in phenomenological diffusion models} \cite{Strong:2007nh}.  Neither provide a good representation of the radial distributions, nor can they convey the inherent asymmetries (as exhibited in Fig.~\ref{fig:star}).  \textit{The enhanced densities near the source position, resembling the expectation from diffusion in one dimension, and concentrations in flux as compared to isotropic propagation lead us to describe these as ``streams''.}

{\bf A short road to death.---}
We have seen, as in Fig.~\ref{fig:sky}, that the appearance of streams remains pronounced as the particle energy is decreased.  At sub-TeV energies, simulations become prohibitive, due to the decreasing $e^\pm$ Larmor radius and the associated need for smaller time steps to accurately compute trajectories to determine the amount of time elapsed before propagation is isotropized.  However, scattering arises from field variations comparable to $r_L$, so for a power-law field spectrum (as used here), a scaling should exist.  Ref.~\cite{Giacinti:2012ar} found a scaling for the time of transition to the diffusive regime for protons as
\begin{equation}
  t_d \sim 10^4  \left(\frac{l_{\rm max}}{150\,{\rm pc}}\right)^\beta \left(\frac{{\rm 1000~TeV}}{E}\right)^\gamma
                          \left(\frac{B_{\rm rand}}{4\,{\mu{\rm G}}}\right)^\gamma \, {\rm yr}
  ,
\label{eq:trans}
\end{equation}
where $\beta\!\simeq\!2$ and $\gamma\!=\!0.25\!-\!0.5$ for Kolmogorov (and similarly for other spectra).

As already noted, energy losses are a vital consideration for electrons and positrons, since the energy loss timescale for high-energy $e^\pm$ due to synchrotron radiation and inverse-Compton scattering is only
\begin{equation}
  t_l \sim 2 \times 10^5  \left(\frac{1~{\rm TeV}}{E}\right) \left(  \frac{1.6\,{\rm eV}\,{\rm cm}^{-3}}{\epsilon_B + \epsilon_\gamma} \right)   {\rm yr}
  ,
\label{eq:loss}
\end{equation}
where $\epsilon_B\!=\!B^2/8\pi\!\approx\!0.6\,$eV$\,$cm$^{-3}$ for $B\!=\!5\,\mu$G.  It is apparent that lower energy $e^\pm$ than those considered in our simulations will be of interest.  These do not experience the Klein-Nishina suppression of higher-energy backgrounds, so we include the estimated infrared and optical backgrounds near Earth \cite{Moskalenko:2005ng}, giving $\epsilon_\gamma\!\sim\!1\,$eV~cm$^{-3}$ (neglecting IC cross section variations with energy).  Since the loss rate deceases with age, $e^\pm$ spend most of their time with energies near that at detection, so we use the proton $t_d$ as a proxy.  We see, for $l_{\max}\!=\!150\!-\!250\,$pc and $B_{\rm tot}\!=\!4\!-\!7.5\,\mu$G, that $t_l\!=\!t_d$ for $E_c\approx\!10\!-\!1000\,$GeV.

The implication is that $e^\pm$ with $E\gtrsim\!E_c$ are expected to lose their energy {\it prior to leaving streams and never reach the isotropic diffusive regime}.  This would have profound consequences for observing $e^\pm$ with energies exceeding $E_c$ at Earth, since their densities are then governed by the structure of the local Galactic field.  This is quite different than when dealing with protons, which experience negligible energy loss and arrive from much greater distances after diffusing through many field domains.

{\bf What about $B_{reg}$?---}
Including an overall regular magnetic field elucidates the origin of the behavior seen above and, as previously mentioned, may more appropriately describe the field within the Galaxy's spiral arms (see also \cite{Gaggero:2013rya}).  In Fig.~\ref{fig:reg3}, we show the outcomes from setting $B_{\rm reg}/B_{\rm rand}\!=\!0$, 1, and 5, while keeping $B_{\rm reg}\!+\!B_{\rm rand}\!=\!3\,\mu$G.  We have again used the source positioned at the center of Fig.~\ref{fig:star}, with $E_g$$\,=\,$1~PeV.  It is clear that, \textit{rather than diminish streams, including a regular field accentuates them}.

Results with a dominant random field, as for $B_{\rm rand}\!\sim\!7.5\,\mu$G and $B_{\rm reg}\!\sim\!3\,\mu$G, are close to a pure random case.  For $B_{\rm reg}\!\approx\!B_{\rm rand}$, streams are aligned with neither the regular field nor the large-scale random components.  In these scenarios, the direction of the nearby ``regular'' field likely lies off of the Galactic plane (as is observed).

For larger $B_{\rm reg}/B_{\rm rand}$, we see that propagation becomes ever more one dimensional, with the stream continuing to greater distances.  We also find that the difference in $D_\parallel$ compared to the phenomenological $D$ vanishes.  Even when we decrease $l_{\max}$ to 50~pc, for which Eqs.~(\ref{eq:trans}) \& (\ref{eq:loss}) imply a very large $E_c$ for a purely random field, we see the same basic behavior for $B_{\rm reg}/B_{\rm rand}\!=\!5$ as in Fig.~\ref{fig:reg3}.  Thus, in this regime the importance of streams would likely persist even at energies well below the nominal $E_c$ inferred from the random component.

{\bf Discussion and conclusions.---}
For electrons and positrons at the energies of current and future interest, we have seen that a consideration of energy losses when propagation is treated as occurring in a turbulent or regular magnetic field alone can \textit{significantly} change the expectations for observations, particularly for nearby sources.  \textit{One consequence is a limitation to the number of contributing sources at Earth}.  This holds when $e^\pm$ propagation in the local Galactic magnetic field can be described using a field coherent over a scale of $\sim\!100\,$pc.  We discuss some of the implications that follow.

\emph{\textbf{Cosmic-ray electrons:}}
As illustrated in Fig.~\ref{fig:star}, even for multiple local sources, at progressively higher energies one would be increasingly fortunate to be located in a stream.  Crossed streams, where multiple sources would be inferred, are relatively rare.  This is radically different than expected in isotropic models, where {\it all} nearby sources contribute to the flux and lead to a spectrum containing numerous features \cite{Grasso:2009ma,Malyshev:2009tw}.  However, the measured electron spectrum has a smooth variation \cite{Fermi:2009zk}.

These structures could be very important at the highest energies for $e^\pm$, where losses are severe, and fluxes may be present even at TeV--PeV energies.  Nearby pulsars known to accelerate $e^\pm$ to these energies \cite{Kistler:2009wm} include Geminga \cite{Abdo:2009ku} and Vela \cite{Aharonian:2006xx}; however, since streams need not follow straight paths from the source (see Figs.~\ref{fig:star} \& \ref{fig:reg3}), \textit{enhancements could appear in any direction, even when the source position is known}.  High-energy electron and positron anisotropies would be more complex than the simple dipole oriented with the source of diffusive prescriptions (see, e.g., \cite{Grasso:2009ma,Ackermann:2010ip,Linden:2013mqa}), although not necessarily with larger amplitude for $r_L\!\ll\!l_w$ (the stream width) since particles would still scatter within the streams.

\emph{\textbf{Positrons (and dark matter):}}
Aside from the general rise associated with the positron excess \cite{Adriani:2008zr,Aguilar:2013qda}, positron data are also rather featureless and have been interpreted in terms of a single pulsar.  Indeed, the smaller number of positron sources would imply fewer streams.  One consequence could be an enhanced flux from an otherwise unremarkable source, easing concerns of exceeding the spin-down power of a lone pulsar.  Limiting the sources that can contribute at Earth in this way could account for the smooth $e^-$ and $e^+$ spectra, since $e^\pm$ fluxes from a source outside of a stream are diminished.

The closer propagation is to isotropy, the more likely a particular source contributes a finite flux.  We have seen in Fig.~\ref{fig:star} that such ``clouds'' are occasionally realized.  One can determine volume filling factors as a function of energy, distance, and time.  However, one could argue that the positron excess suggests that at least one source does reach Earth, so we leave such probabilistic arguments to elsewhere to keep focus on the larger points.

However, a scenario in which \textit{none} of the local astrophysical positron sources actually reach Earth becomes a distinct possibility.  At first glance, this would seem to require the diffuse injection of positrons, as from the smooth Galactic dark matter halo.  However, positron models based on annihilations in dark matter substructure (e.g., \cite{Kistler:2009xf,Vincent:2010kv,Slatyer:2011kg,Blanchet:2012vq}) are affected, since each is a continuously emitting source.  While it is even less likely for the largest ``clumps'' to contribute, the subhalo mass function implies numerous less massive objects \cite{Diemand:2008in,Springel:2008cc}.  Any of these could yield an apparent flux larger than expected from isotropic propagation (as assumed in, e.g., \cite{Brun:2009aj,Kuhlen:2009is}). 

\emph{\textbf{Probing our Galactic neighborhood:}}
The complex nearby environment of Earth \cite{Frisch(2011)} could be uniquely probed by $e^\pm$.  AMS recently revealed a very smooth positron fraction with a possible change in slope at $\gtrsim\!100\,$GeV \cite{Aguilar:2013qda}.  While this is tempting to simply ascribe to a change in the source spectrum, this need not be the case, especially since $E_c$ depends strongly on the actual parameters, which vary throughout the Galaxy.

The above simulations utilize a synthetic turbulent field without feedback.  As discussed above, scattering depends on the field at the scale of $r_L$, which leads to an average $D(E)$ varying as a power law (e.g., \cite{Strong:2007nh}).  If the scattering rate for a given energy was greatly increased beyond this, propagation could approach isotropy sooner.
An alternative to investigate is MHD turbulence.  However, it has been shown via simulation that the anisotropy inherent to the MHD cascade leads to even less effective scattering at small scales unless isotropic fast modes are present \cite{Beresnyak:2010yq}, although these appear to be subdominant in the local interstellar medium (ISM) \cite{Haverkorn:2013eqa}.

Cosmic rays themselves can induce waves of order $r_L$ if a gradient exists within the background of cosmic-ray protons \cite{Kulsrud:1969zz,Wentzel:1969}, which can scatter $e^+$ having the same $r_L$.  The energy at which this may become important would be a third scale to compare with Eqs.~(\ref{eq:trans}) \& (\ref{eq:loss}).  This is notoriously dependent upon damping within the ISM, e.g., due to the scattering of ions with neutral atoms that damps waves.  A canonical value for the energy below which such waves are undamped is $\sim\!100\,$GeV \cite{Kulsrud:1971,Cesarsky:1980pm,Farmer:2003mz,Beresnyak:2008ng}, which can be higher or lower (e.g., $\sim\!20\,$GeV for hot ISM in \cite{Yan:2004aq}).
This could be more relevant in the vicinity of sources \cite{Kawanaka:2010uj,Yan:2011kp,Malkov:2012qd}, where the cosmic-ray density is above average.  Simply comparing the wave growth rate estimated in \cite{Giacinti:2013wbp} for streams with $\sim\!10^{48}\,$erg of $e^\pm$ (as for Geminga in \cite{Yuksel:2008rf}) to the above timescales suggests this may become relevant near $\sim\!100\,$GeV.  However, this neglects damping, which can itself be relevant even in regions where H is ionized because He can remain neutral \cite{Spangler:2010nu}, and reduce this significantly.

While the test-particle approach used here is not suited for examining collective effects, we have attempted to initiate the process of addressing the overall issues presented above.  It is hoped that dynamic simulations, likely within the context of MHD fields \cite{Beresnyak:2013ria}, may soon begin to unravel the complexities of the propagation of electrons and positrons at a more detailed level.

Curiously, a difference between electrons and positrons could arise if $e^+$ result from lower-luminosity sources, such as pulsars, and $e^-$ primarily from supernova remnants.  In this case, we would expect a transition from more isotropic to stream-like topologies to occur at an energy depending on the particle spectra of each source, altering expectations for measured spectra and angular anisotropies.  Changes due to propagation alone should thus be kept in mind when interpreting data, such as from AMS.

Lastly, there is manifest relevance for photon emission from $e^\pm$ near sources (largely independent of background gas density, unlike gamma-ray emission from protons).  While we defer more detailed examination to elsewhere \cite{KistlerX}, simplifications arise at the extremely-high $e^\pm$ energies required to produce $\gtrsim\,$TeV gamma rays, since the corresponding relatively-low $e^\pm$ density diminishes collective effects.  \textit{Both the morphology and amplitude of emission could be greatly altered from standard expectations, based on the orientation and ``speed'' of bulk propagation in streams determined by local magnetic field conditions}.

Drawing guidance from Fig.~\ref{fig:star}, we could expect emission ranging from elongated filamentary structures to, if looking ``down the barrel'' of a stream, a fairly-isotropic signal with exceptionally-low implied lateral diffusion.  Surface-brightness-limited searches may even be biased towards such detections, with numerous unidentified TeV sources likely being PWNe \cite{Yuksel:2008rf}.  Enhanced spatial resolution of nearby sources, such as the TeV gamma-ray ``halo'' \cite{Yuksel:2008rf} observed around Geminga \cite{Abdo:2009ku}, by upcoming experiments, such as HAWC \cite{DeYoung:2012mj}, could thus indirectly uniquely inform us about our local ISM.\\

\acknowledgments
We thank Andrey Beresnyak, Brenda Dingus, Chris Fryer, Phil Kronberg, and Todor Stanev for discussions and comments.
MDK acknowledges support provided by NASA through the Einstein Fellowship Program, grant PF0-110074,
AF was supported by the DOE Office of Science and the LANL LDRD Program, and HY by the LANL LDRD Program.
%


\end{document}